\documentclass{elsarticle}
\usepackage{graphicx}
\usepackage{bm} 
\usepackage{amssymb}
\begin{document}
\begin{frontmatter}
\title {Measurement of the response of a liquid scintillation detector to monoenergetic electrons and neutrons}
\author[a]{P. C. Rout }
\author[b]{A. Gandhi} 
\author[c]{T. Basak}
\author[a]{R. G. Thomas}
\author[d]{C. Ghosh} 
\author[a]{A. Mitra} 
\author[a]{G. Mishra}
\author[a]{S. P. Behera}
\author[a]{R. Kujur}
\author[a]{E. T. Mirgule} 
\author[a]{B. K. Nayak}
\author[a]{A. Saxena}
\author[e]{Suresh Kumar}
\author[f]{V. M. Datar}

\address[a]{Nuclear Physics Division, Bhabha Atomic Research Centre, Mumbai-400085,~India}
\address[b]{Amity Institute of Applied Science, Amity University, Noida, Uttar Pradesh-201313,~India}
\address[c]{UM-DAE Centre for Excellence in Basic Sciences, Mumbai University, Mumbai - 400098,~India}
\address[d]{DNAP, Tata Institute of Fundamental Research, Mumbai-400005,~India}
\address[e]{Ex-Nuclear Physics Division, Bhabha Atomic Research Centre, Mumbai-400085,~India}
\address[f]{INO Cell, Tata Institute of Fundamental Research, Mumbai-400005,~India}

\begin{abstract}
The response of the liquid scintillator (EJ-301 equivalent to NE-213) to the monoenergetic electrons produced in Compton scattered $\gamma$-ray tagging has been carried out for various radioactive $\gamma$-ray sources. The measured electron response is found to be linear up to $\sim$4~MeVee and the resolution of the liquid scintillator at 1~MeVee is observed to be
$\sim$~11\%. The pulse shape discrimination and pulse height response  of the liquid scintillator for neutrons has been measured using $^7$Li(p,n$_1$)$^7$Be*(0.429~MeV) reaction. Non linear response to mono-energetic neutrons for the liquid scintillator is observed at E$_n$=5.3, 9.0 and 12.7~MeV. The measured response of the liquid scintillator for electrons and neutrons have been compared with Geant4 simulation.
\end{abstract}
\end{frontmatter}
\section{Introduction}
The measurement of neutron emission in a nuclear reaction is an important probe for the study of basic nuclear reaction dynamics and also its application in nuclear energy and nuclear astrophysics. An array of liquid scintillator (LS) consisting of 60 detectors is being planned to be set up at the  Pelletron-LINAC Facility (PLF), Mumbai, for the measurement of fast neutron spectra. This will be used to measure neutron cross section as low as $\sim$~1~$\mu$b/sr and will complement the existing array of plastic scintillators \cite{pcr1}. The slow component of the scintillation light allows the discrimination of neutrons and $\gamma$ rays on the basis of pulse shape discrimination (PSD) whereas the fast component is used for time of flight (TOF) measurements. This setup will be used for the study of washing out of shell effect~\cite{pcr2} and damping of rotational enhancement of nuclear level density with excitation energies, the measurement of the prompt fission neutron spectrum in the fast neutron induced fission of actinides and  neutron multiplicity for the study of fusion-fission dynamics.  \\
\begin{figure}
\begin{center}
\includegraphics[scale=0.5]{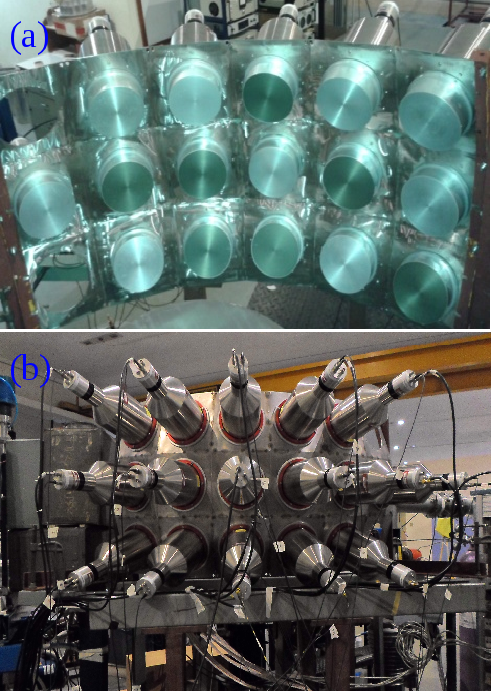}
\caption{Photograph of mini array of liquid scintillator, (a) front view shows the liquid cells and (b) rare view shows the PMTs coupled to the individual liquid cells}
\label{fig01}
\end{center}
\end{figure}
A mini array of 18 liquid scintillators (LS) as a demonstrator, has been set up for fast neutron  spectroscopy by pulse shape discrimination and neutron time of flight technique (TOF). Each of the LS  is cylindrical shape (equivalent to NE213  and procured from SCIONIX, Holland)~\cite{scionix} and has a dimension 12.7~cm diameter and 5~cm thickness. The scintillator has a light output of $\sim$~78\% compared to that of anthracene, a scintillation decay time of $\sim$3.2~ns~(short component) and a bulk attenuation length $>$~2.5~m. Each is LS coupled to a fast linear focused, 12.7~cm diameter Hamamatsu R1250 (14 stage) photo-multiplier tube (PMT) for the signal readout. The detector with a special designed window ensures 100~\% Hermetic seal and can be used in all orientations. The scintillator has a carbon to hydrogen ratio of $\sim$1:1.2 and refractive index is 1.5. The spectral sensitivity of the PMT peaks at 420~nm, with a quantum efficiency of $\sim$ 22~\%, and matches the emission spectrum of the liquid scintillator. The PMTs have a fast response time (rise time $\sim$1.3~ns) and a gain of $\sim$10$^7$ at about 2~kV bias voltage. The PMTs are powered by a 48 channel programmable high voltage power supply developed in-house similar to Ref.~\cite{manna}. These scintillators were mounted on  a mechanical stand, placed  at a flight path$\sim$75cm with an angular separation among the detectors is ~16$^\circ$. A photograph of the detector setup is shown in the Fig.~\ref{fig01}. Along with the TOF the pulse shape discrimination property of the LS is exploited for unambiguous identification of neutrons from the gamma rays. The signals are processed using analogue electronics and stored using a multi-parameter VME based data acquisition system~\cite{lamps}. \\
The paper is organized in four sections. The first section gives details of measurement carried out using radioactive sources for the study of the response to electrons. The next section describes TOF, PSD and pulse height response to mono energetic neutrons measured using the $^7$Li(p,n$_1$)$^7$Be*~(0.429 MeV) reaction at proton energies from 8 to 16~MeV. The subsequent section describes the Geant4 simulation of the LS response and followed by summary.

\section{Response to electrons}
\begin{figure}
\begin{center}
\includegraphics[scale=1.5]{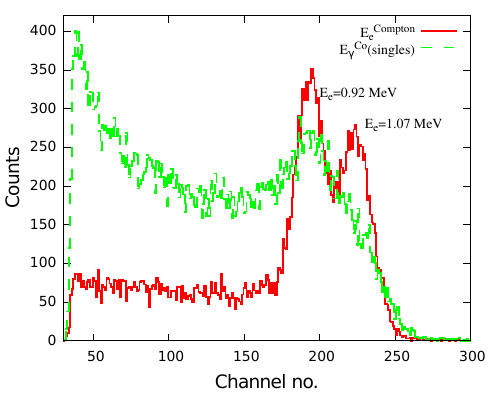}
\caption{Comparison between spectra of liquid scintillator using $^{60}$Co source for measurements without and with coincidence with scattered gamma rays}
\label{fig02}
\end{center}
\end{figure}
Liquid scintillation detectors offer very fast timing information in addition to the pulse shape discrimination for neutron and $\gamma$-rays and thus widely used for neutron spectroscopy. The knowledge of absolute efficiency which is a function of incident neutron energy and threshold is essential for unfolding of the neutron spectra. Accurate determination of the threshold for the analysis of the neutron spectra and neutron response function require very precise energy calibration. The position of pulse height associated with the Compton scattered $\gamma$-ray spectrum introduced the error in the energy calibration. Hence a coincidence between Compton scattered gamma rays and the recoil electron offers a better choice for the energy calibration in the scale of electron energy equivalent~\cite{pcr1,rd}.
\begin{figure}
\begin{center}
\includegraphics[scale=1.0]{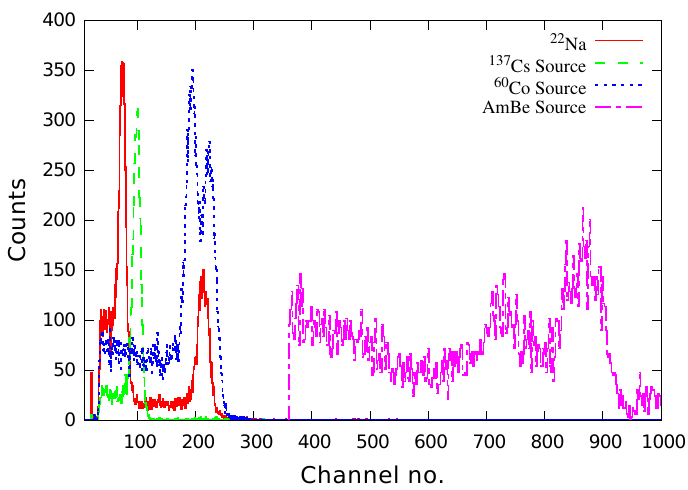}
\caption{ Pulse height response of liquid scintillator up to $\sim$4MeV electron equivalent energy using $^{22}$Na,$^{137}$Cs, $^{60}$Co and $^{241}$Am-$^{9}$Be sources}
\label{fig03}
\end{center}
\end{figure}
\subsection{Experimental Details}
A coincidence experiment has been carried out using a 5" diameter and 2" thick cylindrical liquid scintillator(LS) and a hexagonal, $\sim$6~cm face to face, 8~cm thick BaF$_2$ detector for the study of the LS response to electrons. The LS belongs to the existing mini neutron detector array set up at  the PLF, Mumbai. For the measurements of pulse height response, the $\gamma$-rays were collimated using 2" thick Pb bricks. The Compton scattered $\gamma$-rays from the LS were detected by the BaF$_2$ detector placed at $\theta\sim$135$^\circ$ with respect to the incident direction. The recoil electron energy($E_e$) which deposited in the LS depends on the scattering angle and given by, $E_e=\frac{\alpha E_\gamma(1-cos\theta)}{1+\alpha(1-cos\theta)},$ where $ \alpha=E_\gamma/m_ec^2$ in Compton scattering of $\gamma$-rays. The anode signal from the LS was split into two parts, one fed to the constant fraction discriminator(CFD) for the TOF measurement and the other one sent to the Mesytec MPD4 for the pulse height and the pulse shape discrimination(PSD) information. The anode signal of the BaF$_2$ detector was send to the CFD for generating the start signal to the time analyser for the TOF measurement. The dynode signal from the BaF$_2$ detector was fed to the spectroscopic amplifier for the pulse height information of the scattered gamma rays. The parameters such as energy signal  of the BaF$_2$ detector, the LS energy, PSD and TOF with respect to the BaF$_2$ detector, were recorded in an event-by-event mode using a VME based data acquisition system.  A coincidence between the LS and BaF$_2$ detector was used as the master trigger. 
\begin{figure}
\begin{center}
\includegraphics[scale=1.2]{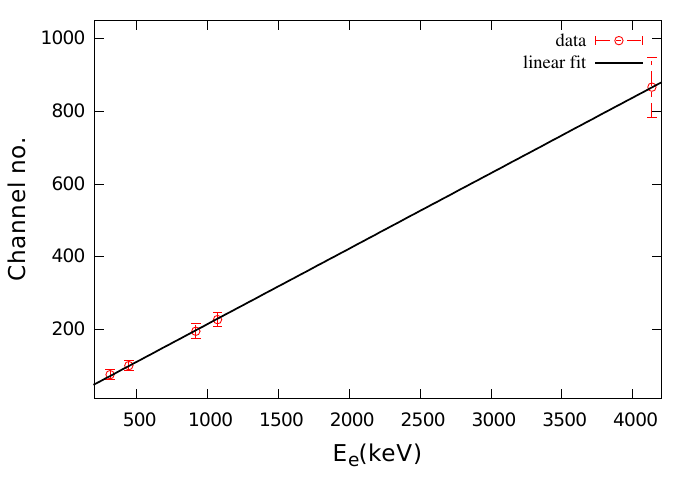}
\caption{ Linearity of light output in electron equivalent energy(E$_e$).}
\label{fig04}
\end{center}
\end{figure}
\subsection{Results and Discussion}
The recoil electron spectrum obtained using $^{60}$Co source tagged with the back scattered gamma rays showed two well resolved peaks with energies 0.9 and 1.1~MeV respectively, while no peak like structure observed for the measurement without coincidence as shown in Figure.\ref{fig02}. This method demonstrated the unambiguous peak definition of the pulse height for the purpose of the calibration of the LS. The resolution of the LS was found to be 11\% at $\sim$1~MeVee~(MeV electron equivalent). The pulse height response tagged by Compton scattered gamma rays was carried out using radioactive sources and the measured response up to $\sim$4~MeVee is shown in Figure.\ref{fig03}. In the present experiment, the measured response of the LS to the electrons was found to be linear up to 4~MeVee as shown in Figure.~\ref{fig04}.
\section{Response to neutrons}
\subsection{Experimental Details}
\begin{figure}
\begin{center}
\includegraphics[scale=0.4,angle=90]{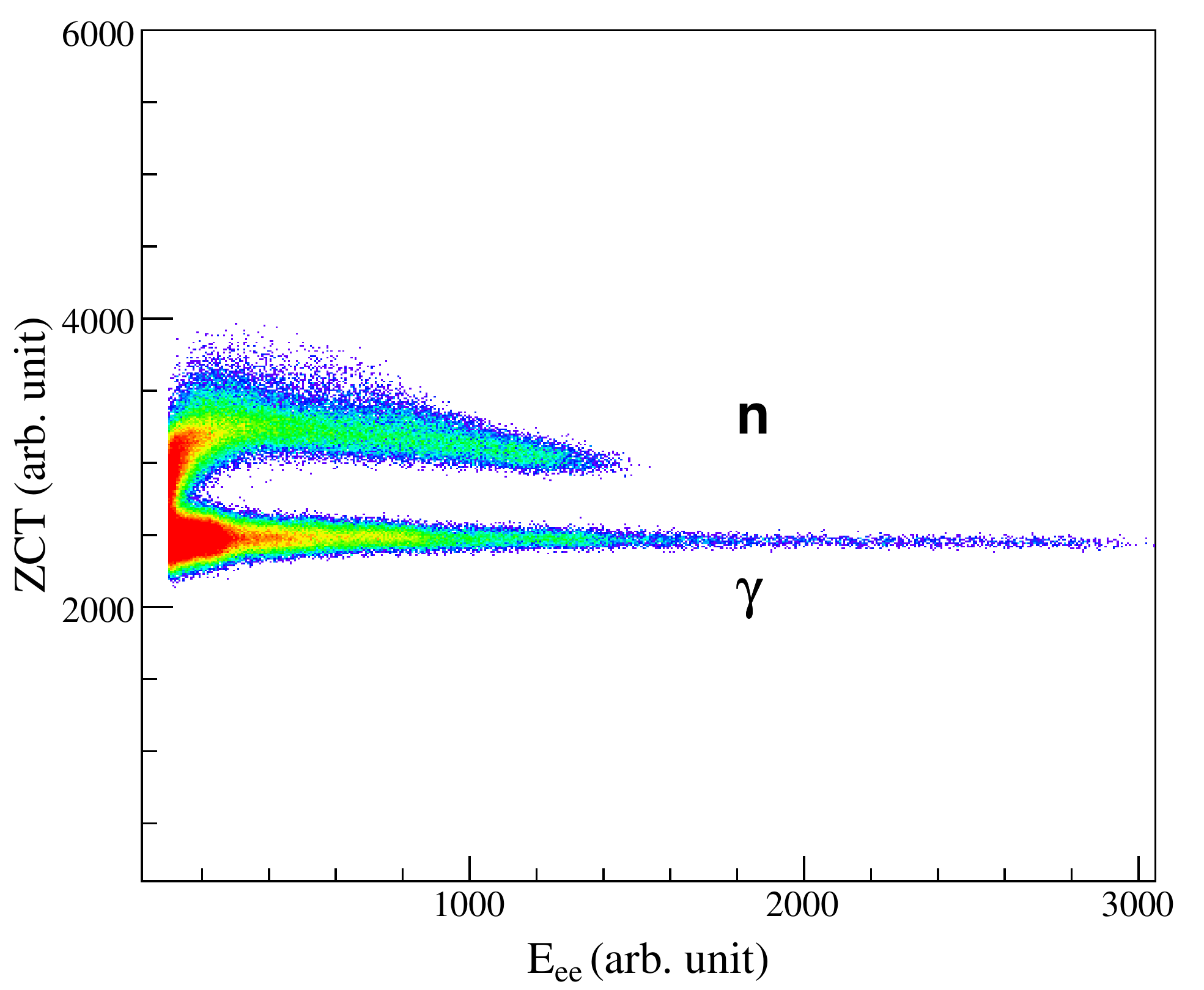}
\caption{A PSD spectra discriminating n and $\gamma$ events for 8~MeV in (p,n$_1$) reaction.}
\label{fig05}
\end{center}
\end{figure}
The experiment was performed to measure mono-energetic neutrons in $^7$Li(p, n)$^7$Be reaction using 8, 12, 16 MeV proton beam at the PLF, Mumbai. The neutrons were measured in coincidence with 429 keV $\gamma$-rays from the first excited state of $^7$Be. A 127 mm diameter and 51 mm thick LS detector (EJ~301 equivalent NE213) coupled to 127 mm diameter photo multiplier tube was used to detect neutrons. The detector was placed at an angle of 45$^\circ$ with respect to the beam direction and at a distance of 1m from the target. An array of seven closed-packed hexagonal BaF$_2$ detectors was placed close ($\sim$3 cm) to the target for detecting the $\gamma$-ray. In each event, the zero cross over time (ZCT), energy deposited in the LS, TOF with respect to the BaF$_2$ array and energies of BaF$_2$ detectors were recorded using a CAMAC based data acquisition system. The TOF was calibrated using a commercial time calibrator. The energy calibration of the LS was performed by measuring the energy of the Compton scattered electrons which were tagged by the back scattered $\gamma$ rays detected in a NaI(Tl) detector as described earlier. 
\begin{figure}
\begin{center}
\includegraphics[scale=0.40]{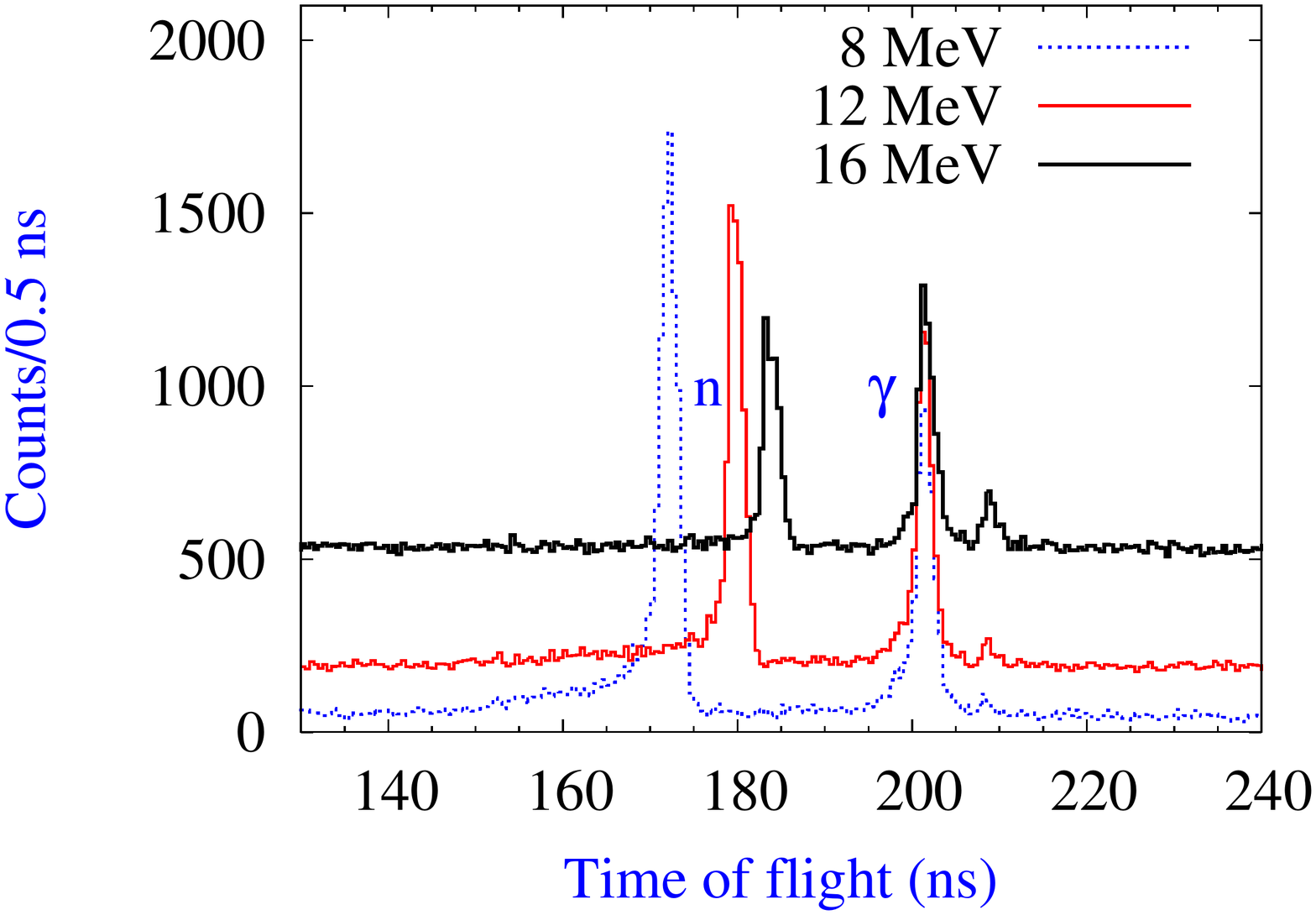}
\caption{Measured neutron TOF spectra at three proton energies.}
\label{fig06}
\end{center}
\end{figure}
\begin{figure}
\begin{center}
\includegraphics[scale=1.4]{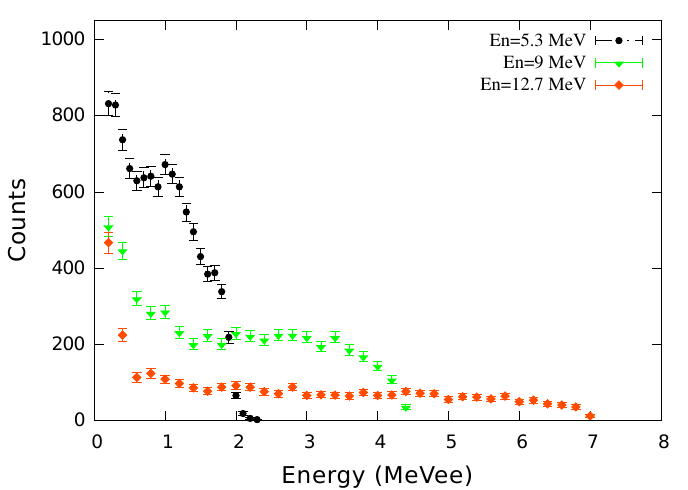}
\caption{Measured Pulse height responses of the liquid scintillator at E$_n$=5.3, 9, 12.7~MeV using (p,n$_1$) reaction.}
\label{fig07}
\end{center}
\end{figure}

\subsection{Results and discussion}
The ZCT  measured with respect to the leading edge timing signal is
a measure of the PSD. Fig.\ref{fig05} shows the measured PSD spectrum exhibiting an excellent neutron and $\gamma$ discrimination for 
E$_{ee}\ge$ 30 keV. The figure of merit (M) for this discrimination is defined as the ratio of the peak separation to 
the sum of their full widths at half maximum for E$_{ee}$=1 MeV. For the present case M~$\sim$~1.8 has been obtained. Neutron TOF spectra measured in coincidence with 429 keV $\gamma$ rays are shown in Fig. \ref{fig06}. The extracted neutron energies from the TOF spectra are tabulated in Table~\ref{T1} and found to be in agreement with the kinematics. 
\begin{table}
\caption{Measured neutron energies~(E$_n^{Meas}$) deduced from the time of flight spectra and calculated electronic equivalent light output~(E$_{ee}$) using the empirical relations~\cite{rac,aksoy} for various proton energy~(E$_p$) in the (p,n$_1$) reaction. E$_n^{Kin}$ is the neutron energy calculated using 2~body kinematics for $^7$Li(p,n$_1$)$^7$Be reaction at $\theta = 45 ^\circ$. E$_{ee}^{end}$ is the end point energy of the measured pulse height spectrum in Fig.~\ref{fig07}. All the figures in the table are in the units of MeV. }
\label{T1}
\begin{center}
\begin{tabular}{|c|c|c|c|c|c|}
\hline
E$_p$ & E$_n^{Kin}$ & E$_n^{Meas}$ &  E$_{ee}^{end}$ &E$_{ee}$~\cite{rac} & E$_{ee}$~\cite{aksoy}\\\hline
8.0& 5.3 & 5.26$\pm$0.25 & 2.3 &2.42$\pm$0.16&1.94$\pm$0.15\\\hline
12.0& 9.0 &8.98$\pm$0.55 &4.4 &5.04$\pm$0.41 &4.23$\pm$0.35\\\hline
16.0& 12.7 &12.67$\pm$0.91& 7.0 &7.89$\pm$0.42 &6.64$\pm$0.61 \\\hline
\end{tabular}
\end{center}
\end{table}
It can be mentioned that the the pulse height response of the liquid scintllator to the neutrons is nonlinear and normally measured using the monoenergetic neutron produced via (p,n) and (d,n) reactions. The empirical relations obtained by comparing the experimental observations~\cite{rac,aksoy} are used to compare the presently measured pulse height response of the liquid scintillator. The empirical relation for obtaining the electron energy equivalent light output over a wide energy range is given by~\cite{rac}, 
\begin{equation}
E_{ee}=a_1T_p-a_2\big[1.0-exp(-a_3{T_p}^{a_4})\big]
\end{equation}
where, the parameters are a$_1$=0.83, a$_2$=2.82, a$_3$=0.25 and a$_4$=0.93. T$_{p}$ is the energy of the recoil nucleus.
Similarly, the empirical relation obtained from (d,n) reaction is given by~\cite{aksoy}  
\begin{equation}
E_{ee}=a_1T_p + a_2T_p^2+a_3T_p^2
\end{equation}
where, the parameters are a$_1$=-1.092, a$_2$=0.5517 and a$_3$=0.00461. \\
The measured pulse height spectrum of LS was extracted after applying suitable prompt and random gates on the  TOF  parameter. The random subtracted pulse height spectra are shown in Fig. \ref{fig07} for various neutron energies. The maximum pulse height equivalent to electron energy calculated using these empirical relations~\cite{rac,aksoy} for the liquid scintillators. The end point energies(E$_{ee}^{end}$) of the measured pulse height spectra reasonably agree with the values calculated using the non-linear response of the LS using these relations are presented in the Table~\ref{T1}. The entire pulse height response can be simulated using the Monte Carlo simulation.
\begin{figure}
\begin{center}
\includegraphics[scale=0.25]{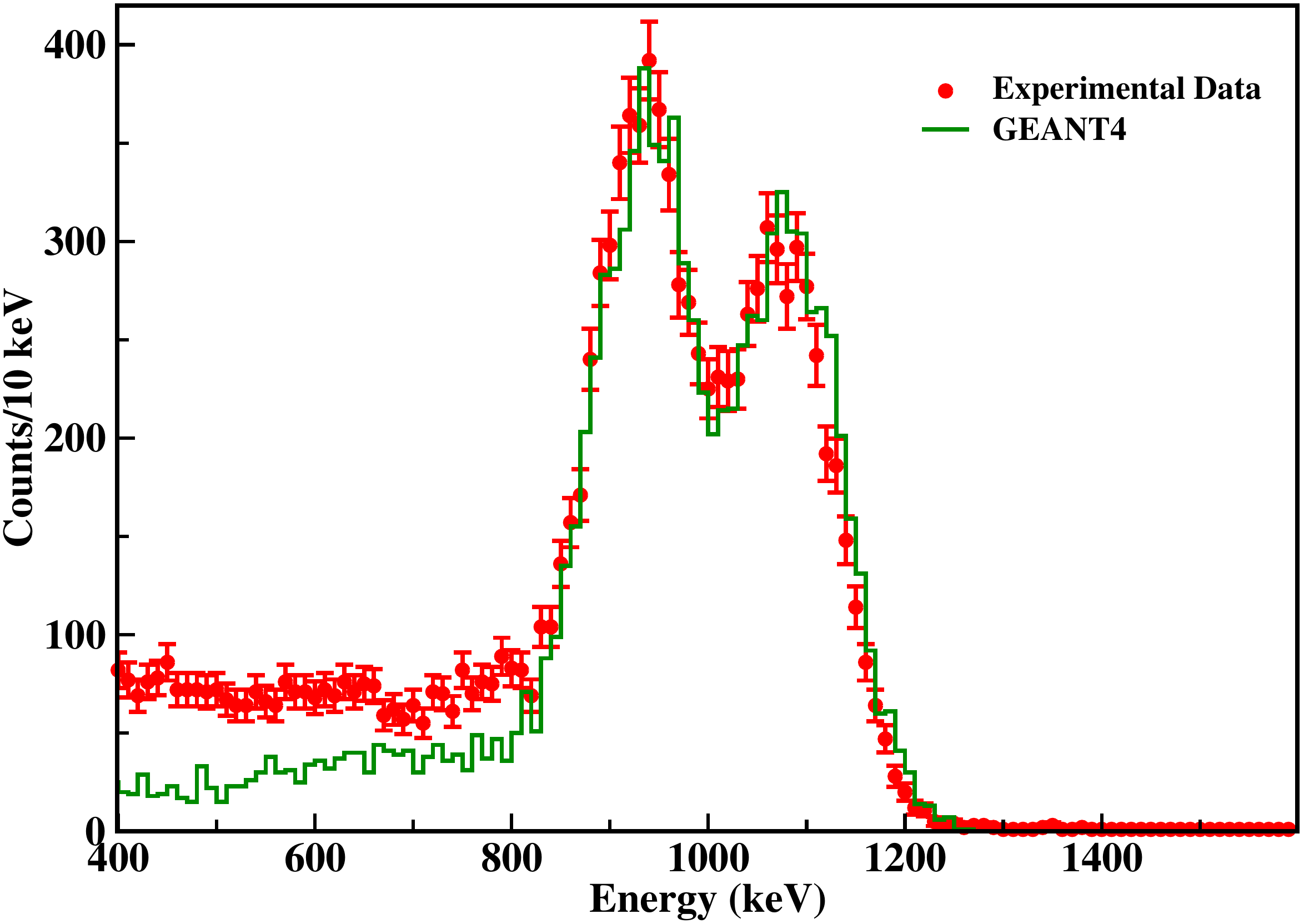}
\caption{A Comparison between the measured and Monte~Carlo simulated electron response using Geant4.}
\label{fig08}
\end{center}
\end{figure}
\section{Geant4 Simulation for the pulse height response of the LS}
\begin{figure}
\begin{center}
\includegraphics[scale=0.3]{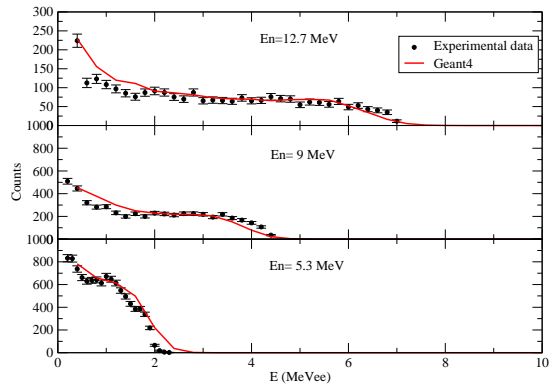}
\caption{Pulse height responses of the liquid scintillator at E$_n$=5.3, 9, 12.7~MeV and solid lines are the Monte Carlo simulation of the energy response using Geant4.}
\label{fig09}
\end{center}
\end{figure}

A Monte Carlo simulation was performed using Geant4 simulation tool-kit~\cite{geant} to study the response of the liquid scintillator. The geomerty and dimension of the LS detector used in the simulation was the same as describe above. The active volume of the detector was taken as 55\% of hydrogen and 45\% of carbon. A typical simulation was performed for 10$^9$ events. The suitable electromagnetic process (photoelectric, Compton scattering , pair production and annihilation, electron Bremsstrahlung, electron scattering and electron ionisation) were used to simulate the electron response of the detector while the neutron cross section data and QGSP\_BIC\_HP physics list were used for the simulation of fast neutron response. The simulated data of LS detector was convoluted with detector resolution of 11.0\% at 1~MeVee and filtered with energy deposition corresponding to Compton scattered events at 135$^\circ$ in BaF$_2$ detector. The filtered spectrum is then compared with the experimental spectrum generated with similar condition. The area under the peak in the LS detector is used to normalized the simulated and experimental data. Figure~\ref{fig08} shows the comparison of simulated and experimental data for $^{60}$Co source and in the peak region the match is reasonably well. The Geant4 simulation for the neutron response using the non-linear response of the liquid scintillators was performed. The simulated response at E$_n$=5.3, 9 and 12.7 MeV are shown in Fig. \ref{fig09} and observed that the simulated neutrons response reasonably explain the measured neutron response for the liquid scintillator.
\section{Summary}
In summary, we report the response of  a liquid scintillation detector (EJ-301) to monoenergetic electrons and  neutron.  The characterization of a representative detector belongs to the mini LS array has been carried out for the PSD, TOF and pulse height response using monoenergetic electrons by measuring Compton tagged electron and monoenergetic  neutrons produced via the $^7$Li(p,n$_1$)$^7$Be*~(0.429 MeV) reactions. The observed electron response is found to be linear up to $\sim$4~MeVee in the present experiment and resolution of the detector at 1~MeVee is about 11.0\%. The figure of merit for the neutron and $\gamma$-ray discrimination is 1.8 at 1~MeV and separation down to 30~KeVee is observed. The measured response of the liquid scintillator to electrons and neutrons agrees well with the response predicted by Geant4 simulation. 

\noindent {\bf Acknowledgements}\\
We thank  Dr. D. R. Chakrabarty for his help and valuable discussion during the experiment. We also thank the target laboratory for Li targets and the PLF staff for smooth operation of the accelerator.

\end{document}